\documentclass{emulateapj}

\usepackage{natbib}

\usepackage{graphicx}
\usepackage{latexsym}
\usepackage{amsfonts,amsmath,amssymb}
\usepackage{url}
\usepackage[utf8]{inputenc}
\usepackage{fancyref}
\usepackage{hyperref}
\hypersetup{colorlinks=false,pdfborder={0 0 0},}

\def\msol{$M_{\odot}$}

\def\lya{Ly$\alpha$}

\shorttitle{Hubble Frontier Fields: The Faint-end of the UV Galaxy Luminosity Function at $z \sim 7-8$}
\shortauthors{Atek et al.}

\begin{document}


\title{New Constraints on the Faint-End of the UV Luminosity Function at $z \sim 7-8$ using the gravitational lensing of the Hubble Frontier Fields Cluster A2744 \footnotemark[$\dagger$]}

\footnotetext[$\dagger$]{Based on observations made with the NASA/ESA Hubble Space Telescope, which is operated by the Association of Universities for Research in Astronomy, Inc., under NASA contract NAS 5-26555. These observations are associated with programs 13495, 11386, 13389, and 11689. STScI is operated by the Association of Universities for Research in Astronomy, Inc. under NASA contract NAS 5-26555. The Hubble Frontier Fields data were obtained from the Mikulski Archive for Space Telescopes (MAST).}

\author{Hakim Atek\altaffilmark{1}}
\author{Johan Richard\altaffilmark{2}}
\author{Jean-Paul Kneib\altaffilmark{1,3}}
\author{Mathilde Jauzac\altaffilmark{4,5}}
\author{Daniel Schaerer\altaffilmark{6,7}}
\author{Benjamin Clement\altaffilmark{2}}
\author{Marceau Limousin\altaffilmark{3}}
\author{Eric Jullo\altaffilmark{3}}
\author{Priyamvada Natarajan\altaffilmark{8}}
\author{Eiichi Egami\altaffilmark{9}}
\author{Harald Ebeling\altaffilmark{10}}

\altaffiltext{1}{Laboratoire d'Astrophysique, Ecole Polytechnique F\'ed\'erale de Lausanne, Observatoire de Sauverny, CH-1290 Versoix, Switzerland}
\altaffiltext{2}{CRAL, Observatoire de Lyon, Universit\'e Lyon 1, 9 Avenue Ch. Andr\'e, 69561 Saint Genis Laval Cedex, France}
\altaffiltext{3}{Aix Marseille Universit\'e, CNRS, LAM (Laboratoire d'Astrophysique de Marseille) UMR 7326, 13388, Marseille, France}
\altaffiltext{4}{Institute for Computational Cosmology, Durham University, South Road, Durham DH1 3LE, U.K}
\altaffiltext{5}{Astrophysics and Cosmology Research Unit, School of Mathematical Sciences, University of KwaZulu-Natal, Durban, 4041 South Africa}
\altaffiltext{6}{Observatoire de Gen\`eve, Universit\'e de Gen\`eve, 51 Ch. des Maillettes, 1290, Versoix, Switzerland} 
\altaffiltext{7}{CNRS, IRAP, 14 Avenue E. Belin, 31400, Toulouse, France}
\altaffiltext{8}{Department of Astronomy, Yale University, 260 Whitney Avenue, New Haven, CT 06511, USA}
\altaffiltext{9}{Steward Observatory, University of Arizona, 933 North Cherry Avenue, Tucson, AZ, 85721, USA}
\altaffiltext{10}{Institute for Astronomy, University of Hawaii, 2680 Woodlawn Drive, Honolulu, Hawaii 96822, USA}

\begin{abstract}

Exploiting the power of gravitational lensing, the Hubble Frontier Fields (HFF) program aims at observing six massive galaxy clusters to explore the distant Universe far beyond the limits of blank field surveys. Using the complete {\em Hubble Space Telescope} observations of the first HFF cluster Abell 2744, we report the detection of 50 galaxy candidates at $z \sim 7$ and eight candidates at $z \sim 8$ in a total survey area of 0.96 arcmin$^{2}$ in the source plane. Three of these galaxies are multiply-imaged by the lensing cluster. Using an updated model of the mass distribution in the cluster we were able to calculate the magnification factor and the effective survey volume for each galaxy in order to compute the ultraviolet galaxy luminosity function at both redshifts 7 and 8. Our new measurements reliably extend the $z \sim 7$ UV LF down to an absolute magnitude of $M_{UV} \sim -15.5$. We find a characteristic magnitude of $M^{\star}_{UV}=-20.63^{+0.69}_{-0.56}$ mag and a faint-end slope $\alpha = -1.88^{+0.17}_{-0.20}$, close to previous determinations in blank fields. We show here for the first time that this slope remains steep down to very faint luminosities of 0.01$L^{\star}$. Although prone to large uncertainties, our results at $z \sim 8$ also seem to confirm a steep faint-end slope below 0.1$L^{\star}$. The HFF program is therefore providing an extremely efficient way to study the faintest galaxy populations at $z > 7$ that would otherwise be inaccessible with current instrumentation. The full sample of six galaxy clusters will provide yet better constraints on the build-up of galaxies at early epochs and their contribution to cosmic reionization.
\vspace{0.3cm}
\end{abstract}


\section{Introduction} 

The study of the luminosity distribution of high-redshift galaxies provides one of the most important clues for understanding galaxy formation and evolution and for investigating the epoch of cosmic reionization. The slope and the normalization of the luminosity function (LF) can be compared to the mass function of dark matter haloes to understand the connection between galaxies and dark matter evolution \citep{behroozi13,birrer14}. This helps to illuminate the relationship between mass and light over cosmic time, a key ingredient in understanding galaxy formation and assembly. Since the galaxy luminosity function in the rest-frame ultraviolet (UV) encapsulates direct information about the efficiency of star formation, this enables measurements of the star formation density and its evolution across cosmic time using the LF \citep{bunker04,bunker10,bouwens11,oesch14}. Finally, the census of star-forming galaxies at $z > 6$ through the UV LF allows us to assess the contribution of galaxies to the ionization history of the intergalactic medium \citep{oesch09,trenti10,castellano10,castellano10b,bouwens12b,finkelstein12,robertson13}.

Great progress has been made during the last decade with the detection of large samples of galaxies out to the highest redshifts due to powerful telescopes and instrumentation \citep[e.g.][]{sawicki06,bouwens07,reddy09}. Most of these samples were assembled by detecting the ultraviolet continuum break in the broadband images of these galaxies \citep{steidel96, giavalisco02}. The Wide Field Camera 3 (WFC3) on board the {\em Hubble Space Telescope} ({\em HST}) has extended these kinds of studies to fainter and more distant galaxies thanks to a higher sensitivity and a larger field of view \citep[e.g.,][]{bunker10,bouwens11,schenker13,mclure13,bouwens14,schmidt14}.

Obtaining a comprehensive picture of cosmic star formation hinges on understanding the physical properties of UV-selected galaxies and the detection of the faintest population that lies beyond the detection limits of current surveys. Such dwarf galaxies may contribute significantly to the total star formation density of the Universe at $z > 2$. Indeed, {\em HST} grism spectroscopy shows that low-mass galaxies at $z \sim 2$ are dominated by violent starbursts and tend to have much higher specific star formation rates (sSFRs) than their massive counterparts \citep{amorin14,maseda14, atek14c,masters14}. There is also evidence pointing to the increase in number of such high-sSFR galaxies towards higher redshifts \citep{shim11, atek11,stark13, debarros14,smit14} and of this population of low-mass galaxies being the likely culprits for cosmic reionization. The abundance of feeble galaxies at $z > 7$, together with a relatively high escape fraction of Lyman continuum, may be enough to sustain the ionization state of the IGM \citep{robertson10,bouwens12b,nestor13}.

One can push the limits of current instrumentation to study very faint high-redshift galaxies by using foreground galaxy clusters as a magnifying glass. The strong gravitational lensing of massive systems amplifies the flux of background sources while spreading the light over a larger area \citep{kneib11}, although there are smaller areas which are ``de-magnified''. Therefore, it is possible to detect much fainter galaxies than what is possible in blank fields \citep{richard12,bouwens12,bradac12,bradley13,balestra13,coe13,monna14,schmidt14b,zheng14,atek14b,karman14,richard14b} and perform spatially detailed studies of these distant galaxies \citep{jones10,zitrin11,frye12,sharon12,kawamata14,jones14}. The galaxy cluster A1689 was used in \citet{alavi14} to probe the UV-LF at $z \sim 2$ down to very faint magnitudes ($M_{1500} = -13$ AB mag). One of the important findings of that study is the absence of turnover in the faint-end of the LF at such low luminosity. The contribution of fainter galaxies of course results in a higher star formation rate density than previously determined.

The {\em Hubble} Frontier Fields (HFF) is a large DDT (Director Discretionary Time) program that exploits the power of ``gravitational telescopes'' to probe the distant Universe to unprecedented depth by targeting six massive clusters. A total of 140 orbits are devoted to each cluster, split between four WFC3 (Wide Field Camera 3) near-infrared filters and three ACS (Advanced Camera for Surveys) optical filters. The first HFF cluster Abell 2744 (hereafter A2744) was first observed in late 2013. Early results based on the first half of the data clearly demonstrate the ability to detect highly-magnified galaxies, and hence to extend the UV-LF studies to very faint magnitudes \citep{atek14b,laporte14,zheng14,coe14,yue14} thanks to considerable efforts in the modeling of the cluster mass distribution \citep{zitrin13,medezinski13,grillo14,umetsu14,richard14,johnson14,diego14,jauzac14,jauzac14b,mohammed14,montes14,donahue14}. In this paper, we investigate the UV-LF at redshift $z \sim 7$ and 8 based on the full optical and IR observations of A2744. The outline of the paper is as follows. We describe the observations and data reduction process in Section \ref{sec:obs}, catalog construction in Section \ref{sec:catalog}, and the cluster lens modeling in Section \ref{sec:models}. The galaxy dropout selection procedure is presented in Section \ref{sec:selection} while the candidates are presented in Section \ref{sec:candidates}. We show our UV-LF results in Section \ref{sec:LF}, and conclude in Section \ref{sec:conclusion}. We use a standard $\Lambda$CDM cosmology with $H_0=71$\ km s$^{-1}$\ Mpc$^{-1}$, $\Omega_{\Lambda}=0.73,$\ and $\Omega_{m}=0.27$. Magnitudes are in AB system.  

\vspace{1cm}

\section{HFF Observations}
\label{sec:obs}

These {\em HST} observations are part of the HFF program (GO/DD 13495) and include NIR and optical data centered on the cluster A2744 at RA$=$00:14:21.2, DEC$=-$30:23:50.1. NIR imaging includes the four bands F105W, F125W, F140W, and F160W, and was obtained during October and November 2013 for a total of 24, 12, 10, and 24 orbits, respectively. The second epoch observations of A2744 started in May 2014 to obtain ACS imaging in F435W, F606W, and F814W filters for a total of 18, 10, and 42 orbits, respectively. The observational details are summarized in Table \ref{tab:obs}. The reduced and calibrated mosaics used here were obtained from the HFF science data products release 1.0\footnote{http://www.stsci.edu/hst/campaigns/frontier-fields/FF-Data}. Basic reductions were handled by the Space Telescope Science Institute (STScI) who also performed additional calibrations before the final combination, such as a correction for the sky background variations in the NIR images and charge transfer inefficiency residuals in the optical frames. The individual frames were then combined using {\tt Astrodrizzle} with a final pixel scale of 0.06\arcsec pix$^{-1}$.    

\begin{table}
\caption{\label{tab:obs} Summary of {\em HST} observations}
\begin{tabular}{lccc}
 Instrument/Filter & \# Orbits & Depth \footnote{The depth of the images are 3-$\sigma$ magnitude limits measured in a 0.4\arcsec aperture.} & Obs Date \\ \hline 
  WFC3/F160W  & 24  & 28.3&Oct/Nov 2013\\
 WFC3/F140W  & 10  & 29.1&Oct/Nov 2013  \\
 WFC3/F125W  & 12  & 28.6&Oct/Nov 2013 \\
 WFC3/F105W  & 24  & 28.6&Oct/Nov 2013 \\
 ACS/F814W & 18  & 29.4 &Jun/Jul 2014\\ 
 ACS/F606W & 10 & 29.4&Jun/Jul 2014 \\ 
 ACS/F435W & 42  & 28.8 &Jun/Jul 2014 \\ \hline
\vspace{0.3cm}
\end{tabular}
\end{table}

\section{Photometric Catalogs}  
\label{sec:catalog}

Using a point spread function (PSF) model derived with {\tt Tiny Tim} \citep{Krist_Hook_Stoehr_2011}, we first matched all the images to same PSF of the F160W image. Then we created a deep IR image by combining the four IR frames weighted by the inverse variance map (IVM). Similarly, we created a deep optical image combining the F435W and F606W images and a deep ACS image combining the three optical filters in order to ensure that the $z \sim 7$ and 8 candidates are not detected in the blue bands (cf. Sect. \ref{sec:selection}). We used the same procedure as in \citet{atek14b} for source extraction. Using the SExtractor software \citep{bertin96} we performed the source detection in the deep image and measured the photometry in each of the bands in an isophotal aperture (ISO). The ISO magnitude is used for color-color selection. The total magnitude is measured in a Kron radius. We used RMS weighting maps based on the IVM derived during the drizzling process. The background was estimated locally for each object during the flux measurement. Finally the flux errors were corrected for the correlated noise resulting from the drizzling method following \citet{r_Hook_Levay_Lucas_et_al__2000}.

\vspace{1cm}
\section{Strong Lensing Model}
\label{sec:models}

We present here the main aspects of the strong lensing mass modeling of A2744. A detailed description can be found in \citet{jauzac14} where the new model based on the most recent observations of A2744 are presented. Following the approach of \citet{natarajan97} the mass contribution of bright massive cluster galaxies are also used to build the basic model \citep{limousin07, richard14}. The large scale distribution of dark matter consists of four halos around the locations of four bright cluster members. In addition to this large scale distribution, we include perturbations induced by galaxy-scale halos from cluster members using scaling laws based on their magnitude \citep{limousin07}.

 The optimization of the mass model is performed in the image plane using the {\tt Lenstool} software \citep{jullo07}. Using this model we predict the positions of the multiply-imaged systems, and do so by identifying or confirming candidates selected by color and morphology. We identify 50 background systems that provide a total of 152 multiple images. Four of these systems have spectroscopic confirmation, adding to the reliability of the model. The best fit model is found after an iterative process using the locations of the multiple images. The model presented in Jauzac et al. is a significant improvement over previous versions based on pre-HFF observations \citep{johnson14,richard14}. Here for this work, we use the new model to compute the lensing magnification map and to generate the critical lines at different redshifts.

\section{Dropout Selection}
\label{sec:selection}

We select high-redshift candidates using a color-color diagram to detect the continuum break caused by the IGM absorption and to minimize contamination by very red low-z interlopers. This method was first used to select Lyman break galaxies (LBGs) at $z > 2$ with a very high success rate \citep[see][for a review]{giavalisco04}. To sift out the redshift $z \sim 7$ dropout candidates, we adopted the following color criteria:

\begin{align}
\label{eq:criteria7}
 (I_{814} {-} Y_{105})   &>  0.8  \notag \\
(I_{814} {-} Y_{105})   &>  0.6 + 2(Y_{105} {-} J_{125})\\
(Y_{105} {-} J_{125})  &< 0.8 \notag
\end{align}

These color cuts are similar to what has been used in \citet{atek14b} but are slightly different from what has been used in previous studies \citep[e.g.,][]{oesch10} with the aim of including more galaxies at $z > 7$ and fewer galaxies near $z \sim 5$. In order to avoid the contamination from red galaxies that are at $z \sim 1$, we discard any galaxy that shows a significant detection (at a 2-$\sigma$ level) in the $B_{435}$ and $V_{606}$) bands or the $B_{435} + V_{606}$ stacked image. In case an object was not detected in the $I_{814}$ band we assigned a 2-$\sigma$ detection limit for its magnitude when estimating the $I_{814} {-} Y_{105}$ color. We also require that our candidates are detected in the $Y_{105}$ and $J_{125}$ bands with a minimum of 5-$\sigma$ significance. 

To select $z \sim 8$ candidates, we use the following criteria: 

\begin{align}
\label{eq:criteria8}
(Y_{105} {-} J_{125})  &>  0.5 \notag \\
(Y_{105} {-} J_{125})  &>  0.4 + 1.6(J_{125} {-} H_{140})\\
(J_{125} {-} H_{140})  &< 0.5 \notag
\end{align}

\begin{figure*}[tb]
\begin{center}
\includegraphics[width=18cm]{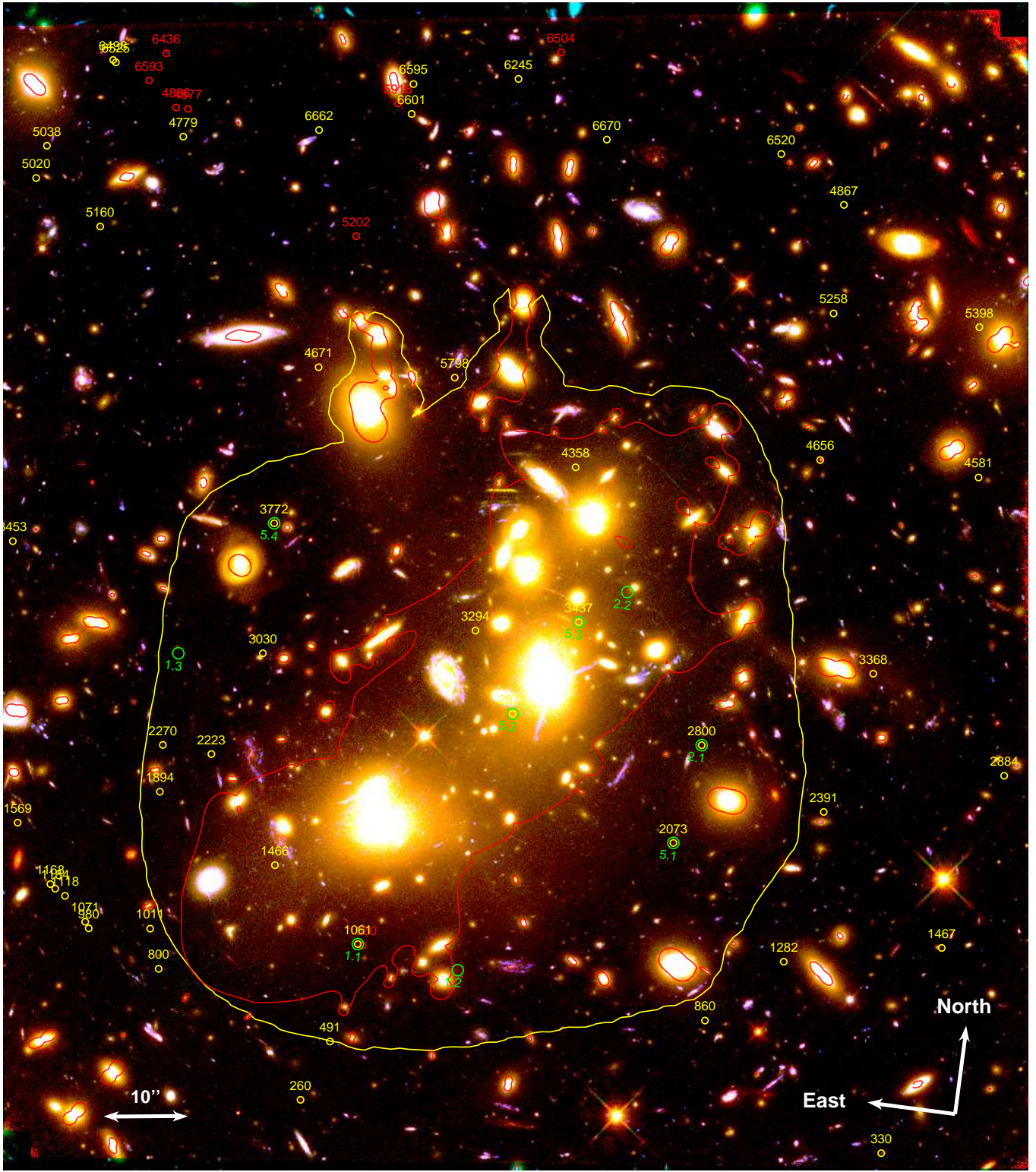}
\caption{\label{fig:layout} Color image of A2744 showing the strong lensing model and the position of the dropout candidates. The pseudo-color image consists of a combination of F435W and F606W filters in the blue channel, F814W in the green channel, and a deep stack of the four IR filters in the red channel, respectively. The yellow circles denote the position of the $z \sim 7$ candidates while the red circles show the position of the $z \sim 8$ candidates. We also show the position of the multiple images of the three identified systems marked with green circles. The red curve is the critical line at redshift $z =7$ derived from the lens model. The yellow curve delimits the region where multiple-image systems are expected.}
\end{center}
\end{figure*}

\begin{figure*}[tb]
\begin{center}
\includegraphics[width=9cm]{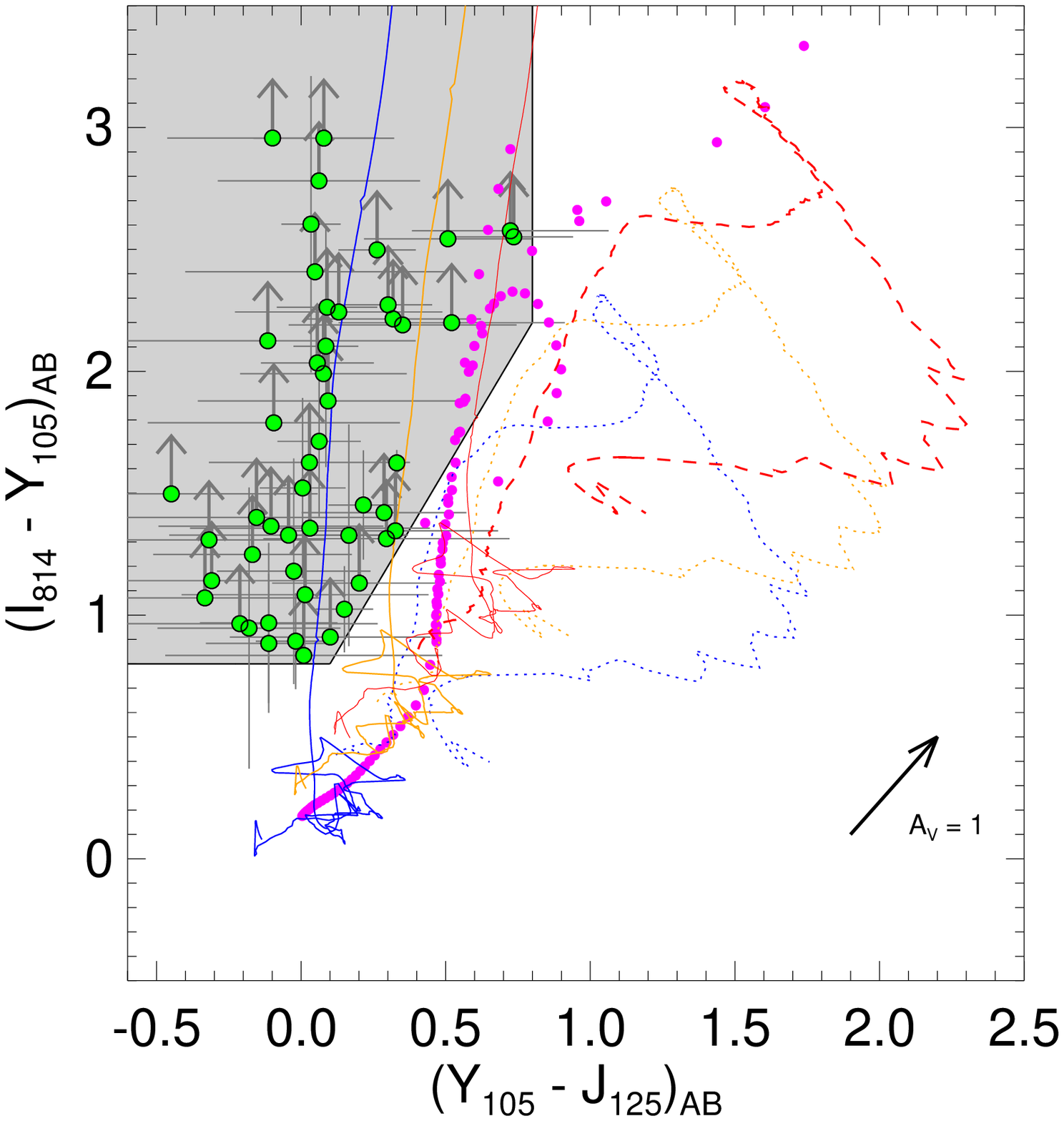} \hspace{-0.5cm}
\includegraphics[width=9cm]{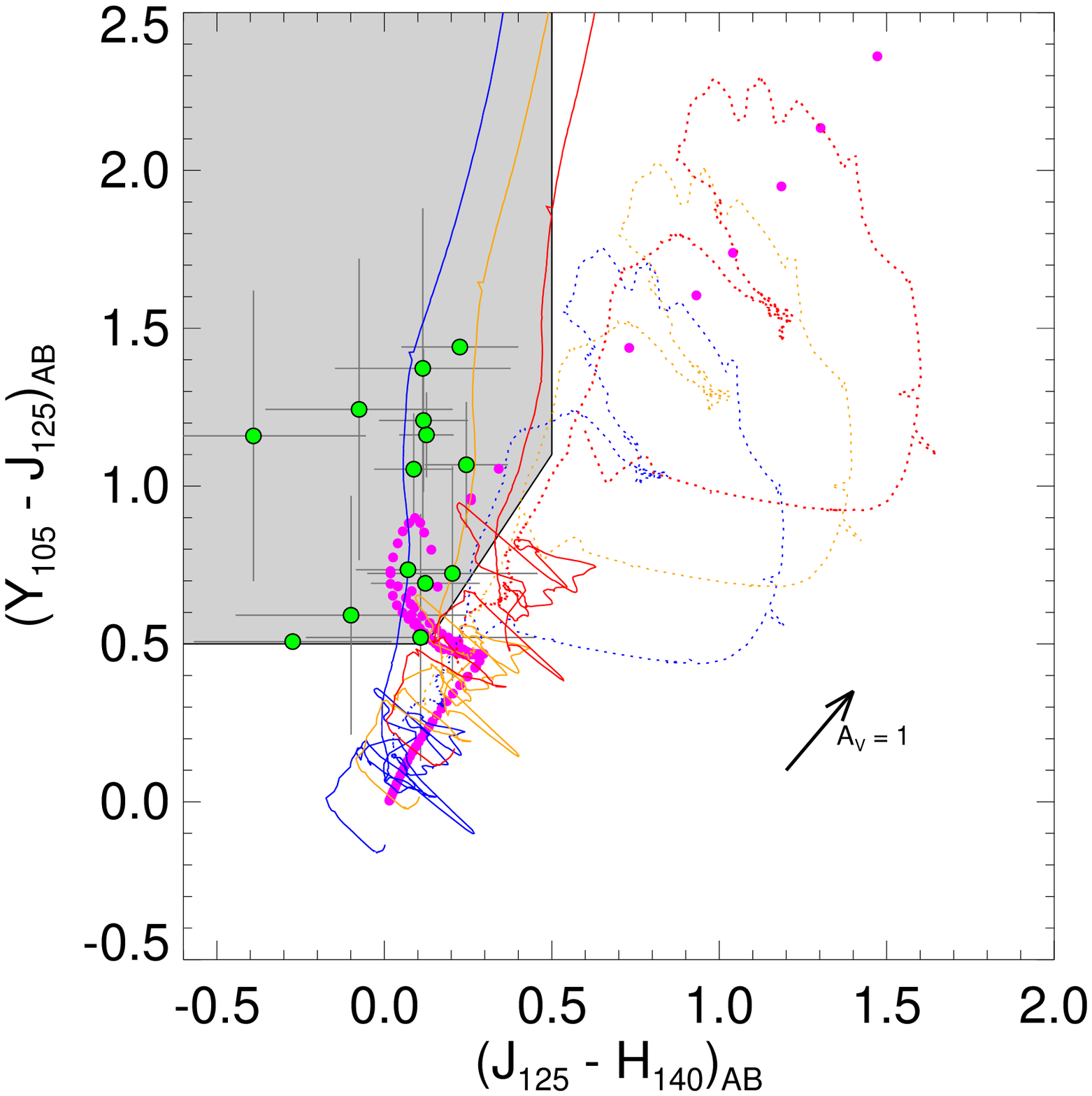}\hspace{-0.5cm}
\caption{\label{fig:color_selection} Color-color diagrams used to select high-redshift galaxies at $z \sim 7$ (left panel) and $z \sim 8$ (right panel). The selected candidates are represented by green circles within the selection window (shaded region). The color tracks of elliptical galaxy templates of \citet{coleman80} are shown with dotted lines while starburst galaxies from \citet{kinney96} are shown in solid lines.Three values of attenuation are applied for both templates: $A_{v}$=0 (blue), $A_{v}$=1 (orange), and $A_{v}$=2 (red). The black arrow shows the direction of the shift caused by the dust extinction. The magenta points show the color evolution of stars \citep{chabrier00} within the diagram.}
\end{center}
\end{figure*}

Similarly, we require no detection in the deep ACS image for $z \sim 8$ candidates. These deep optical images have a detection limit at least one magnitude deeper than the object magnitude in the detection band. In Fig. \ref{fig:color_selection} we present the color-color selection at $z \sim 7$ and $z \sim 8$. The $I_{814}$ and $Y_{105}$-dropouts are marked with green circles while the color tracks represent different types of galaxies as a function of redshift. Elliptical galaxy templates (dotted lines) are based on stellar libraries of \citet{coleman80} and star-forming galaxies (solid lines) on templates from \citet{kinney96}. Three assumed values for the attenuation are represented in blue ($A_{V}$=0), orange ($A_{V}$=1), and red ($A_{V}$=2). We also computed the expected colors of cool stars and those are shown in magenta. The shaded region shows our selection window optimized to minimize contamination from these low-redshift objects. Additionally all candidates were visually inspected to identify spurious detections, such as diffraction spikes, and apertures affected by blending and contamination issues.

\section{High-redshift Candidates}
\label{sec:candidates}

With the completion of the {HST} observations of A2744, we use optical images much deeper than the ancillary ACS data used in \citet{atek14b} for instance, which enables us to select even fainter candidates. In \citet{atek14b} we restricted our search to objects brighter than $J_{125} \sim$ 28 mag to ensure the detection of the break in the $I_{814}$ filter which had a 1-$\sigma$ limiting magnitude of about 29. The depth of the new ACS observations reaches 29.8 AB at 2-$\sigma$ level, allowing the detection of objects as faint as $J_{125} \sim$ 29. We detect a total of 50 candidates at $z \sim 6-7$ with observed total magnitudes ranging from $J_{125} \sim 24$ to $J_{125} \sim 29$ AB magnitudes. This sample includes half of the candidates selected in \citet{atek14b}. The other half did not satisfy our modified criteria to select higher-redshift galaxies. Few galaxies appear brighter in the $I_{814}$ band of the deep ACS observations, leading to a weaker continuum break than observed in \citet{atek14b}. The new break puts these galaxies at a slightly lower redshift than reported in our previous estimate. For few galaxies, this is due to a faint detection in the $V_{606}$ band in the new images. However, the deep optical imaging confirms that these galaxies are at redshifts higher than $z \sim 5$. \citet[][Ze14 hereafter]{zheng14} also published a list of $z \geq 7$ candidates based on the first epoch of IR observations of A2744 and the shallow ancillary ACS data. We detect three out of their 18 candidates in our $z \sim 7$ selection. We note that their redshift selection favors $z > 7.5$ objects because of a different color-color selection window, which shows a larger overlap with our sample at $z \sim 8$.

At higher redshift, we report the detection of eight objects at $z \sim 8$. Three sources in this sample were present in the $7-8$ selection of Ze14, and four others were in their $z=8-9$ category. With this full cycle of A2744 observations, we also confirm the high-redshift candidate at $z \sim 8$ (ID 6436) reported in \citet{atek14b} and discussed in more detail in \citet{laporte14}. For comparison, we also extend our search to $z \sim 9$ galaxies\footnote{For $z \sim 9$ candidates, we used the following color criteria
\begin{align}
\label{eq:criteria9}
(Y_{105} {+} J_{125})/2  &>  0.7 \notag \\
(Y_{105} {+} H_{140})  &>  0.4 +4(H_{140} {-} H_{160})\notag \\
(H_{140} {-} H_{160})  &< 0.7 \notag
\end{align}}. The resulting sample includes eight galaxies already selected in the $z \sim 8$ window. Three more candidates from Ze14 are confirmed by our selection. Overall, one of their robust candidates (ZD8) and all of their potential candidates were not confirmed by our selection. Furthermore, \citet[][Zi14, hereafter]{zitrin14} recently reported the detection of a multiply-imaged system in A2744 with an estimated redshift of $z \sim 10$. We confirm one of the multiple images where object 2994 in our sample corresponds to image B of the $z \sim 10$ galaxy. The primary image A was not selected because of its red color, even though it satisfies the continuum break condition. Our candidate has also a red continuum, falling at the edge of the selection window. The high-redshift solution derived from our photometric redshift estimate is less convincing than what has been reported in Zi14.

\subsection{Sample Contamination}

Among the important sources of contamination are low-redshift evolved galaxies that are heavily attenuated and/or show a strong Balmer break \citep[e.g.][]{boone11,Pello_Schaerer_Le_Borgne_2012}. While these sources can satisfy the dropout criterion they usually show very red colors in bands red-ward of the break. Therefore, most of these objects would be rejected by our third color criterion. Also, our condition on the non-detection in the optical bands, with a detection limit of $\sim 1$ mag deeper than the object magnitude in the IR bands, greatly minimize this source of contamination. The continuum break can also be mimicked by strong nebular lines that contribute to the flux of one of the broadband filters. We do not expect such contamination to be important because, in most cases, we would be able to identify such contribution as an excess in only one or two amongst the four IR filters. Indeed, the relatively smooth continua, mostly flat or blue, in our candidates indicates that strong line emitters are not a concern in our sample. Moreover, as demonstrated by \citet{atek11}, given the depth of the ACS images, most of such interlopers would be detected in the optical bands where the continuum is not affected by emission lines

Another possible source of contamination in our sample are transient objects. The ACS observations were taken several months after the WFC3 ones which can lead to a detection of a supernova explosion for instance in the IR filter which fades out later in the optical bands leading to a fake continuum break. We compared the four epochs of IR data taken over a five week period to check the consistency of the photometry and identify bright transients. However the most robust check rely on the stellarity parameter computed by SExtractor combined with a visual inspection to determine whether the candidates look like point sources. All candidates have a stellarity parameter lower than 0.8 and do not look like point sources in the high resolution {\em HST} images shows. Object 6436 is the most uncertain as, its stellarity is about 0.8 and we can not exclude the possibility of unresolved source. From Figure \ref{fig:color_selection}, we can see that the tracks of brown dwarfs can enter our selection window. Similarly, these can be identified as point sources based on a visual inspection and with their stellarity parameter higher than 0.8 from SExtractor. Finally, the robust detection above 5-$\sigma$ significance of all the candidates minimizes spurious detections resulting from artifacts or faint sources due to photometric scatter. 
 
\begin{figure*}[!htbp]
   \centering
   \includegraphics[width=6.5cm]{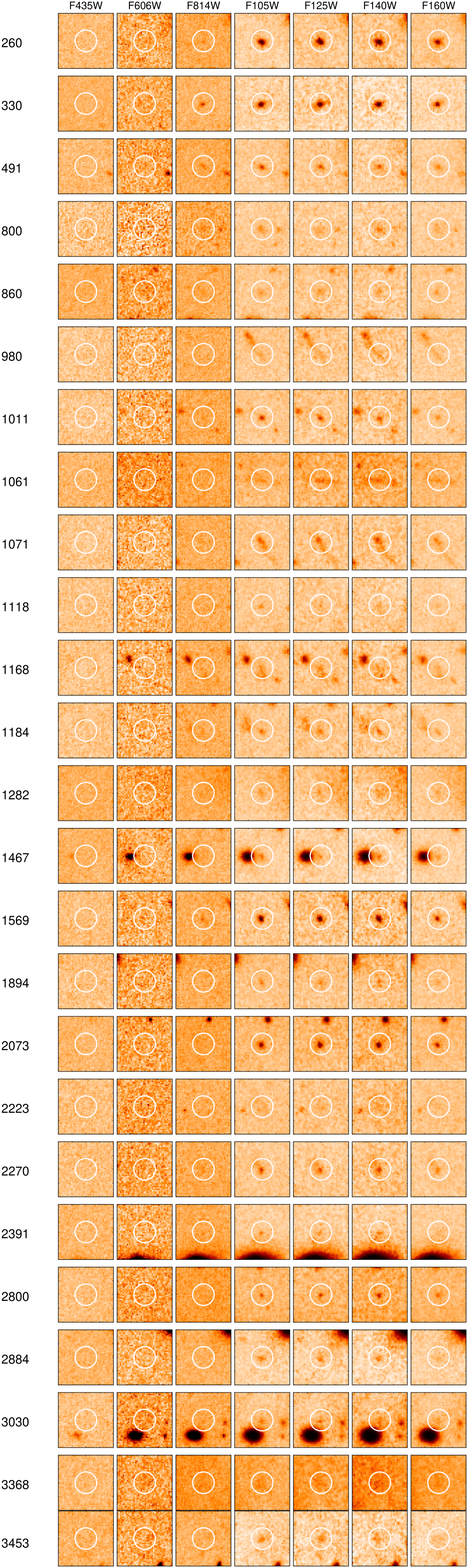} 
    \includegraphics[width=6.5cm]{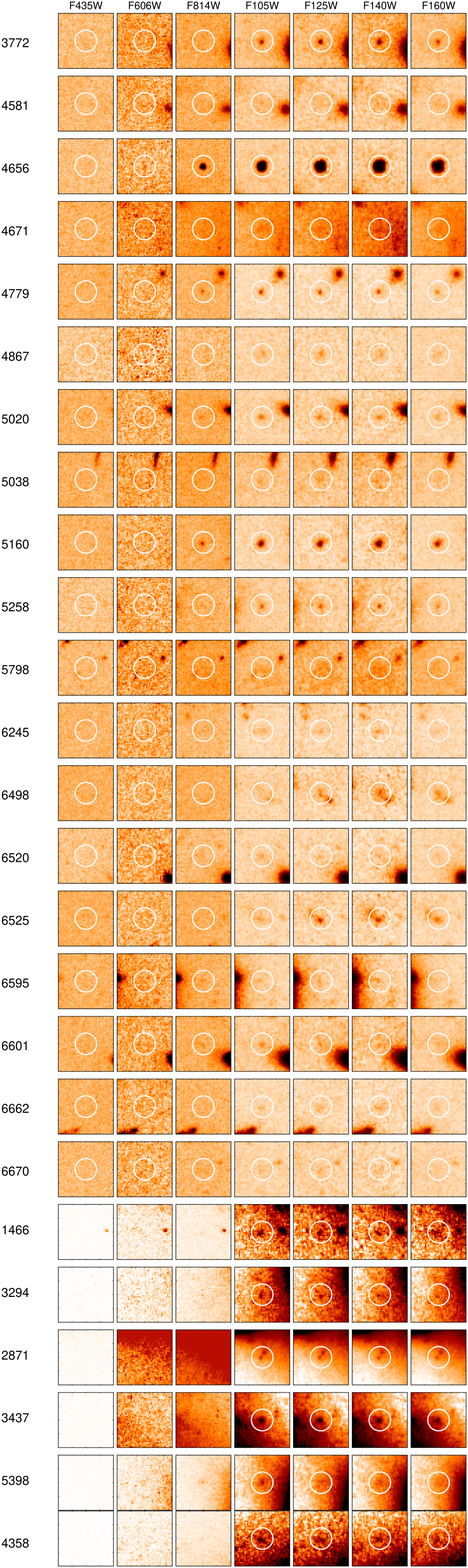} 
   \caption{Image cutouts showing the dropout candidates at $z \sim 7$ with their identification number. Three ACS and four IR images are shown for each candidate with a size of 2.5\arcsec. The white circle denotes the position of each galaxy. }
   \label{fig:cutout_z7}
\end{figure*}

\begin{figure}[!htbp]
   \centering
   \includegraphics[width=6.5cm]{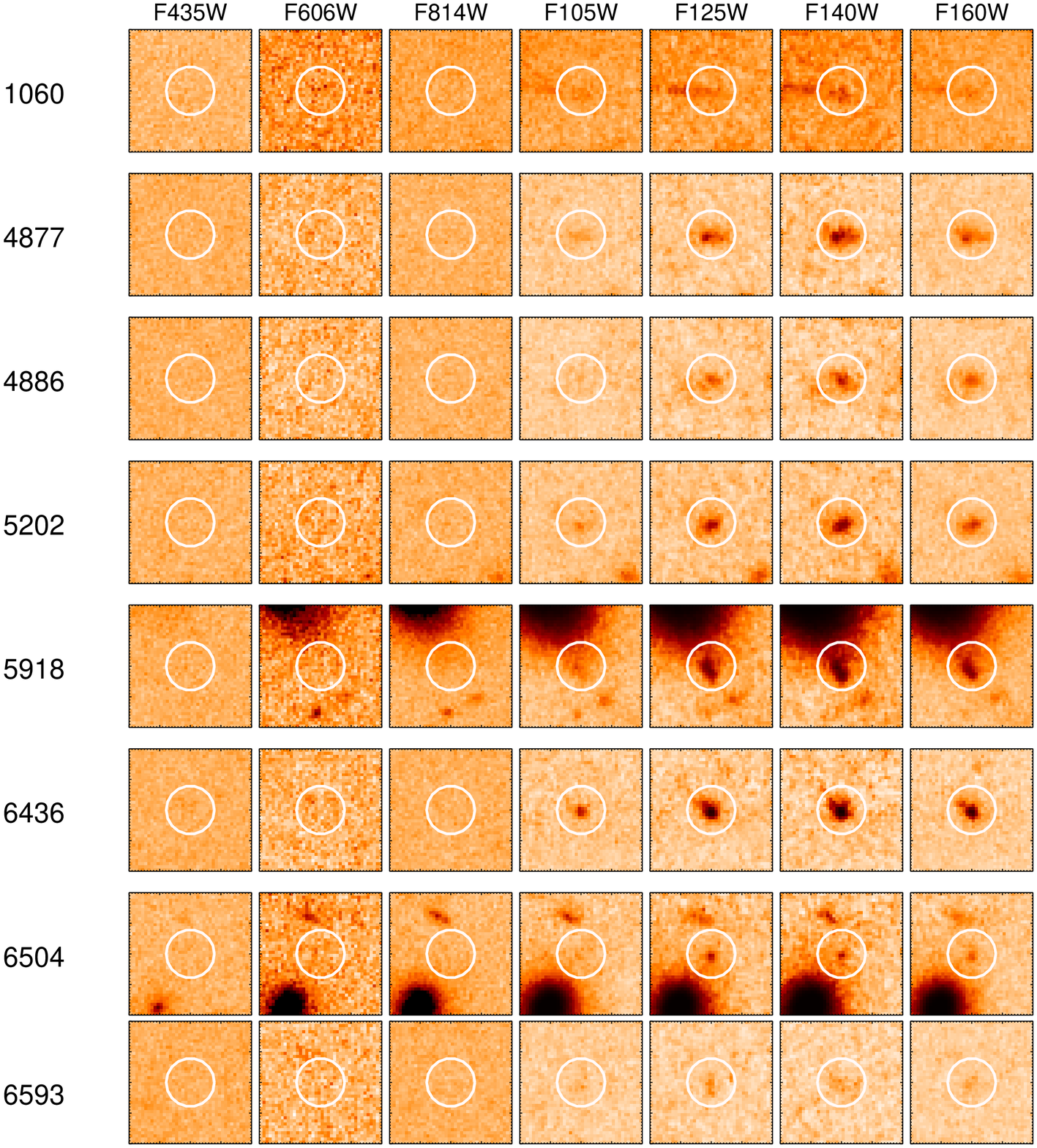} 
   \caption{Same as Figure \ref{fig:cutout_z7} showing postage stamps of our $z \sim 8$ candidates.}
   \label{fig:cutout_z8}
\end{figure}

\subsection{Photometric Redshifts}
\label{sec:photoz}
In addition to our color-color selection, we performed a spectral energy distribution (SED) fitting to the photometry of our candidates to estimate their photometric redshifts. We used a modified version of the {\em Hyperz} code \citep{bolzonella00} that is described in detail in \citet{schaerer09}. This code takes into account the effects of both nebular continuum and emission lines in the fitting of galaxy SEDs \citep{schaerer09, smit14}. We used stellar population libraries from \citet{bc03} with variable star formation histories parametrized by SFR $\propto \exp(-t/\tau)$ with  $\tau=$ (0.01, 0.03, 0.05, 0.07, 0.1, 0.3, 0.5, 0.7, 1., 3., $\infty$) Gyr, a \citet{salpeter55} initial mass function (IMF) from 0.1 to 100 \msol, and metallicities in the range 0.05--1 $Z_{\odot}$.  
The output of the fitting procedure provides the main physical properties of the stellar population and the photometric redshift. For the vast majority of sources analyzed here, the photometric redshifts differ by less than 0.1 with or without the treatment of nebular emission and allowing for \lya\ emission between case B strength (depending on the intrinsic Lyman continuum emission) and zero (to mimic a strong IGM). For all of our candidates the photometric redshift confirms the high-redshift solution. The photometric redshift of each candidate was in turn used to compute its amplification factor using the cluster mass model and {\tt Lenstool}.

\subsection{Multiple-image Systems}

\begin{figure}[!htbp]
   \centering
   \includegraphics[width=7.5cm]{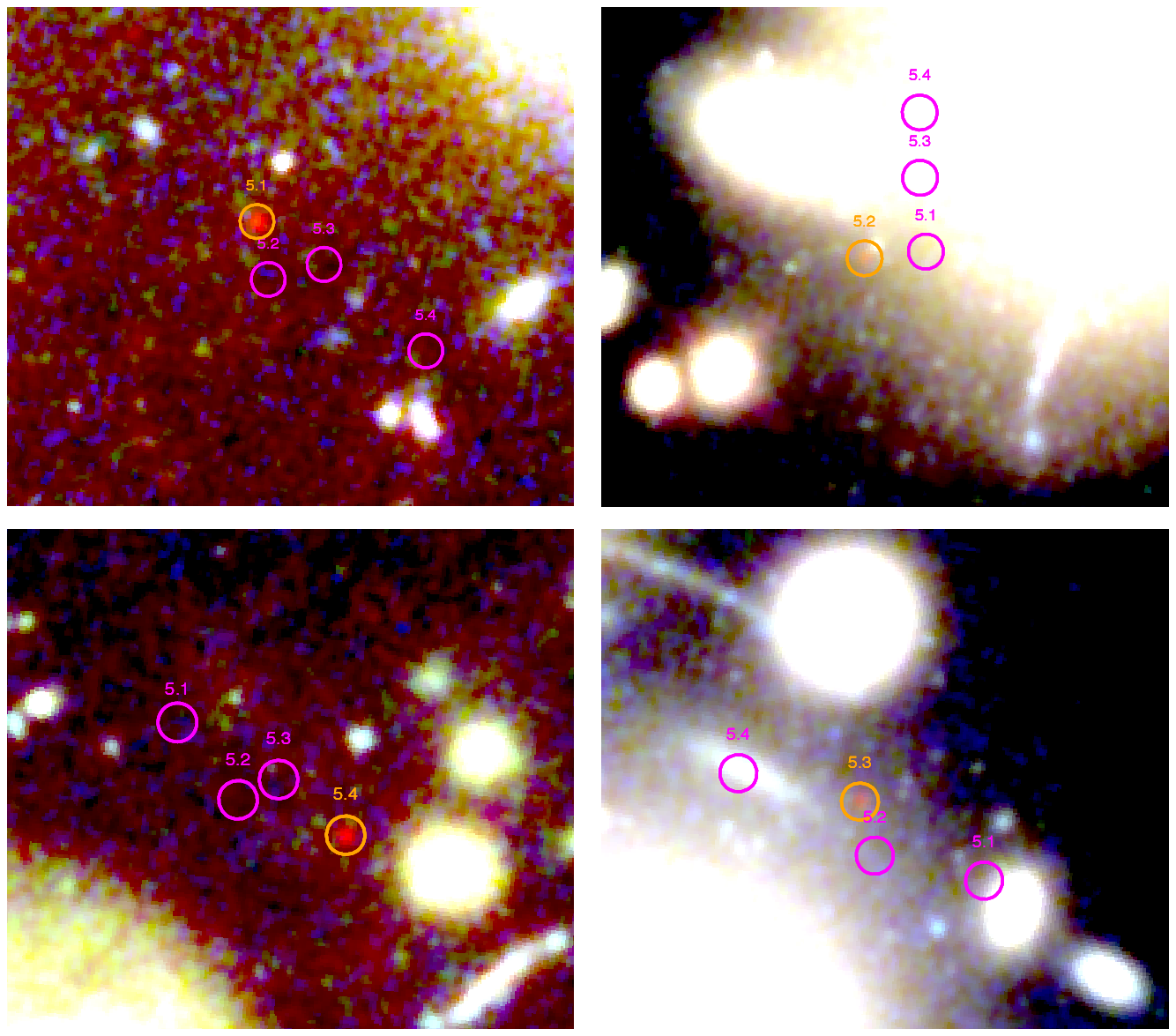} 
   \caption{Multiple image system 5 with four dropout candidates identified at $z \sim 7$. The orange circle in each cutout image denotes the position of the image while the magenta circles show the lens model prediction of the position based on the counter-images. The color image is a combination of F606W (green) F814W (blue) and a stack of four WFC3 IR filters (red).}
   \label{fig:multi}
\end{figure}

Using our updated lens model for A2744 we looked for counter images for each of our strongly lensed candidates. In each case, we determine the position of the corresponding multiple images by ``de-lensing'' the candidate position back to the source plane and reconstructing the full set of images by then ``re-lensing'' the source using {\tt Lenstool}. We then verify the colors and the morphology in the vicinity of the predicted position to find plausible images. The yellow curve in Fig. \ref{fig:layout} denotes the region where we expect multiple images for the same background galaxy to land. A total of 14 candidates in the redshift range probed in this paper fall in this region. The full set of multiple-image ``systems'' at all redshifts is presented in \citet{jauzac14}, which were used as constraints for deriving the new cluster mass model. Table \ref{tab:multiples} summarizes the photometric properties of the multiply-imaged systems we identified. There is one system (main source ID 2073) with four images located around the center of the cluster (see Fig. \ref{fig:multi}). All images were selected as high-$z$ dropouts (cf. Table \ref{tab:multiples}). This system was already identified in \citet{atek14b} where only two images were included in the dropout selection. We corrected the position of one image (5.4) that was previously misidentified. Two other systems (IDs 1061, 2800) exhibit two multiple images, the former of which was independently identified by \citet{lam14}. These counter-images are too faint to be selected by our dropout criteria and are less secure than in system 5 due to large uncertainties in their photometry.

\section{The UV Luminosity Function}
\label{sec:LF}

Using the new sample of high-redshift galaxies found in these new HFF observations, we now construct the UV luminosity function (LF) at $z \gtrsim 6$. We determine the absolute luminosity distribution in bins of absolute magnitude. This is done by applying the lensing amplification correction to the observed flux on the $J_{125}$ band for $z \sim 6-7$ galaxies and to the $H_{140}$ flux for $z \sim 8$ galaxies. The LF is expressed in terms of source volume density per magnitude using the following equation:

\begin{eqnarray}
\phi(M)dM = \frac{N_{i}}{V_{eff}(M_{i})},
\end{eqnarray}

where $N_{i}$ denotes the number of sources in the $i$eth magnitude bin and $V_{eff}(M_{i})$ is the effective survey volume probed at the corresponding bin.

\subsection{Estimating the effective Volume}
\label{sec:volume}

While the main advantage of strong lensing is the amplification of the intrinsic flux of background sources, the downside resides in the reduction of the survey volume due to the stretching of the source plane. Strongly magnified regions will end up probing smaller areas. The advantage of cluster lensing over blank fields will depend on the slope of the UV luminosity function \citep[e.g.][]{maizy10}. A steep bright-end slope yields a positive lensing bias, uncovering a larger number of sources compared to a blank field at a given observed magnitude \citep{richard14,coe14}. While lensing fields lose this numerical advantage towards the faint-end, the magnification nevertheless allows the identification of very faint sources, otherwise inaccessible with current instrumentation. Therefore, lensing effects need to be carefully taken into account while determining the survey volume, and hence the UV luminosity function.

In addition to the lensing bias, we account for a number of incompleteness parameters by performing extensive Monte Carlo simulations. We first generated a sample of 10,000 artificial galaxy SEDs using starburst templates from \citet{kinney96} libraries. Regarding the size of these galaxies, we use a log-normal distribution of the half light radius, which has been shown to be a better representation of the observed sizes in high-redshift LBGs. A uniform distribution includes a higher number of galaxies with relatively large radii which leads to an overestimate of the completeness correction at the faint-end of the LF \citep{grazian11}. \citet{oesch10b} used a similar approach based on a sample of photometrically-selected candidates at $z \sim 7$, with a mean half light radius of 0.153\arcsec and sigma=0.0754\arcsec\ \citep[see also][]{ferguson04}. \citet{grazian11} use the distribution of spectroscopically-confirmed LBGs at $z \sim 4$ presented in \citet{vanzella09} and scale it to high redshift as (1+z)$^{-1}$ following the virial radius evolution for constant halo mass \citep{bouwens04,bouwens06}. They find a similar log-normal distribution at $z \sim 7$ with a mean of 0.177\arcsec and sigma of 0.0751\arcsec. In general, faint galaxies at $z \sim 7$ with $M_{UV} > -20.5$ like all of our candidates, have very small half light radii r$< 0.2$\arcsec\ \citep{grazian12,ono13}. More recently, \citet{kawamata14} report very small sizes of less than 0.2 kpc in faint ($M_{UV} > -19$ AB) $z>6$ galaxies in A2744. they also show that there is no strong correlation between size and luminosity at these faint magnitudes.

  In this paper, we generate random galaxy sizes following a log-normal distribution with a mean of 0.15\arcsec and sigma of 0.07\arcsec. The importance of the size distribution is even more critical for lensed galaxies due to the gravitational distortion. At the faint-end of the LF, the candidates are often close to the detection limit, in which case the lensing stretch of galaxies can lower their significance level below our selection threshold. We also explore the effects of different two-dimensional galaxy profiles on the completeness correction. According to observed morphologies of high-$z$ galaxies in previous studies \citep[e.g.][]{ferguson04,hathi08,grazian11}, we adopt a realistic mix of exponential disks (with a Sersic index of n=1) and de Vaucouleur profiles (with Sersic index n=4). Finally, we use the size-luminosity relation of $r \propto L^{0.25}$ found by \citet{huang13} for high-$z$ LBGs to scale the galaxy size as a function of the intrinsic luminosity.

\begin{figure}[htbp]
   \centering
   \includegraphics[width=7cm]{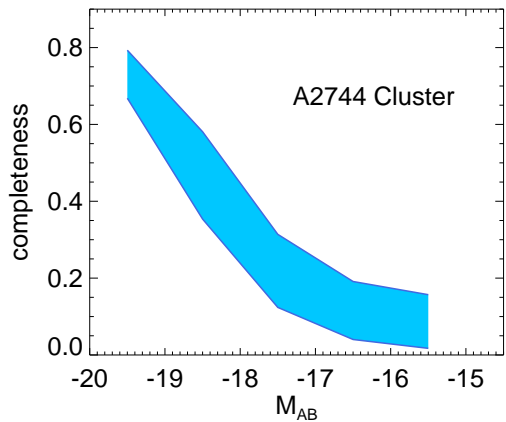} 
   \caption{Survey completeness in detecting high-redshift dropouts in A2744 field. The area represents 68\% confidence intervals computed from Monte Carlo simulations that take into account the lensing effects described in \ref{sec:volume}. Completeness estimate is presented as a function of intrinsic absolute magnitude, i.e. including the amplification factor.}
   \label{fig:comp}
\end{figure}

The next step in the simulations is to assign to the synthetic spectra a random redshift in the range $z=$[5.5,7.5] and an absolute magnitude in the range $M_{UV}$[-24,-14] mag. Galaxies are randomly distributed in the source plane. To simulate lensing effects we lensed the galaxies back to the image plane in the WFC3 field of view. A total of 1,000 real images, including 10 objects in each simulation were created to avoid introducing artificial crowding. Using our magnification map, we derive the amplification at the position of each object. The shape of galaxies is now distorted and their flux amplified according the cluster lens model \citep{jauzac14,richard14}. The apparent magnitudes are then computed by convolving the galaxy SEDs with the throughput curves of the ACS and WFC3 filters. Finally, we run SExtractor on all images to construct photometric catalogs and apply the same color-color selection procedure we applied to select the real candidates from the HFF dataset. The comparison with the input catalogs thus provides the completeness function as function of the different galaxy and lensing parameters described above. In Figure \ref{fig:comp} we present the result of the completeness calculations as a function of the intrinsic absolute magnitude, where the shaded blue region represents 68\% confidence intervals. The completeness level in the faintest magnitude bin goes below 10\% level, accounting for a large contribution to the luminosity function uncertainties.

This completeness function is incorporated in the computation of the effective volume in each magnitude bin following the equation:

\begin{eqnarray}
V_{eff} = \int_{0}^{\infty}  \int_{\mu > \mu_{min}} \frac{dV_{com}}{dz}~ f(z,m,\mu) ~d\Omega(\mu,z) ~dz
\end{eqnarray}

where $\mu_{min}$ is the minimum amplification factor $\mu_{min}$ needed to detect a galaxy with a given apparent magnitude $m$. $f(z,m,\mu)$ is the completeness function computed above, and $d\Omega(\mu)$ is the area element in the source plane as a function of magnification and redshift.

With these effects carefully taken into account, we now present the results of the UV luminosity function at $z \sim 7$ and $z \sim 8$ derived through the lensing cluster A2744. The binning in absolute magnitude is chosen to balance the uncertainty (number of galaxies) in each bin with the sampling of the LF. Naturally, we also exclude multiple images of the same object from the luminosity function calculation. Other sources of uncertainties in the LF determination include Poisson errors and cosmic variance. Very recently, \citet{robertson14} estimated the cosmic variance in high-redshift galaxy samples in A2744, finding a significant increase in the associated uncertainties compared to blank fields. Following the dependance between the survey volume and the lensing magnification they computed the cosmic variance uncertainty as a function of intrinsic magnitude, which we include here in our LF determination.

\subsection{The UV LF at $z\sim 7$}

For redshift $z \sim 7$, we construct five bins at M$_{abs,UV}$=[-19.5,-18.5,-17.5,-16.5,-15.5] with a constant width of 1 mag. We combine our data points with the results of wide area surveys to include better constraints in the fitting of the bright-end of the luminosity function. The small survey area of 0.96 arcmin$^{2}$ probed by the lensing cluster limits the number of relatively bright candidates, thus we discover only galaxies fainter than $M_{abs} \sim -20$. Moreover, for a given detection limit, the number counts are simply shifting to intrinsically fainter magnitudes in lensed fields \citep{coe14}. Using previous estimates of the UV LF at $z \sim 7$, \citet[][in their Figure 15]{richard14} estimated the number of candidates expected in A2744 and other clusters as a function of observed (lensed) magnitude. On average, the total number of 10 galaxies expected at $z \sim 8$ is in agreement with the observed counts we report in this paper and we find slightly more candidates at $z \sim 7$ than expected (about 40 sources). This can be attributed to the redshift selection function \citep[cf.][]{atek14b}, which is broader than what has been used in previous studies or the $\Delta z=1$ used in \citet{richard14}. Cosmic variance from field to field and sample contamination can also of course contribute to the observed difference. Eventually, the full HFF observations of the six galaxy clusters will certainly provide better constraints on the observed number counts.

 \begin{figure}[!htbp]
   \centering
   \includegraphics[width=9cm]{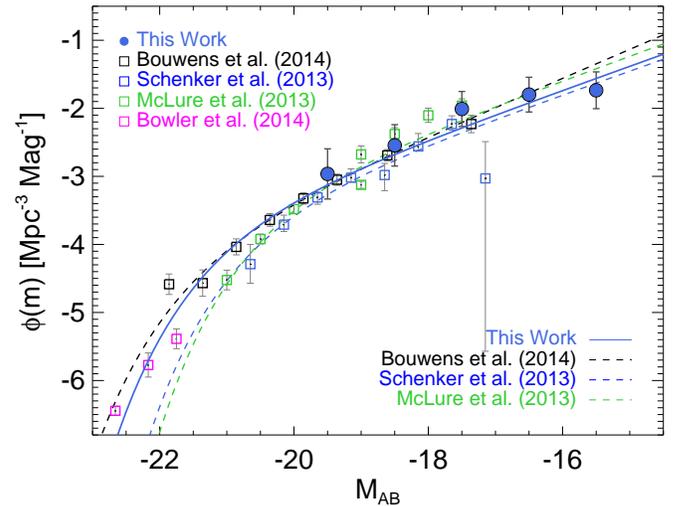} 
     \caption{The UV galaxy luminosity function at $z \sim 7$. Our determination of the intrinsic (unlensed) luminosity distribution based on the effective volume method (see text for the computation details) is shown with the blue circles. The best fit Schechter function is represented with the blue solid line. For comparison, we also show the results that estimate the $z \sim 7$  LF from the literature. Results from a compilation of {\em HST} fields by \citet{bouwens14} are shown in black squares while the dashed black line shows their best fit. Blue squares represent data points from the UDF12 campaign \citep{schenker13} and the dashed blue line represents their best fit. The green squares and the green dashed line represent the results of \citet{mclure13}. We also show data points from wide area surveys of \citet[][magenta squares]{bowler14}.}
   \label{fig:lf7}
\end{figure}

\begin{figure}[!htbp]
   \centering
   \includegraphics[width=8cm]{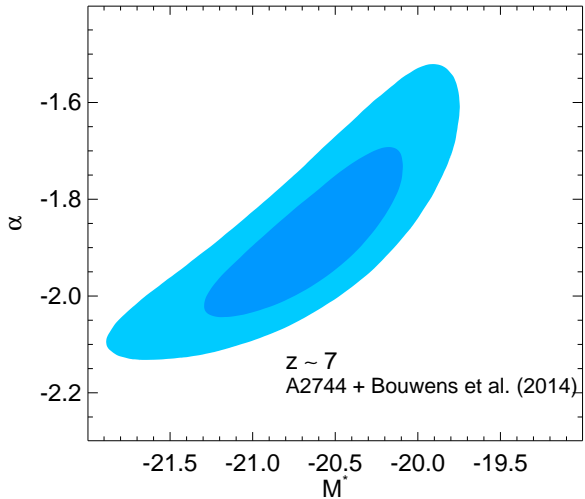} 
     \caption{Confidence intervals inferred from MCMC simulations for the characteristic magnitude $M^{\star}$ and the faint-end slope $\alpha$ of the Schechter fit to the $z \sim 7$ LF. The dark blue and light blue areas show the 68\% and 95\% probability distribution, respectively.}
   \label{fig:lh7}
\end{figure}

In Figure \ref{fig:lf7} our best fit to the the UV luminosity function at redshift $z \sim 7$ is plotted. We performed Markov Chain Monte Carlo (MCMC) simulations to estimate the best fit parameters and the associated uncertainties. For the sake of comparison we show different results from blank fields including deep observations of the ultra deep field 2012 \citep[UDF12,][]{schenker13, mclure13}, wide area surveys \citep{bowler14} and the most recent estimate of \citet{bouwens14} using all available {\em HST} legacy fields. Although we are primarily interested in the faint portion of the luminosity function, there is degeneracy between the characteristic magnitude $M^{\star}$ and the faint-end slope $\alpha$ \citep{bradley12,bouwens14}. However, the additional determinations of the bright-end of $z \sim 7$ LF from wide surveys are still affected by significant uncertainties. While \citet{bowler14} advocate a power-law shape for the bright-end of the UV LF at $z \sim 7$, the results of \citet{bouwens14} are better described by a Schechter function. The disagreement can be observed in Fig. \ref{fig:lf7} around $M_{UV} \sim -22$ mag. We choose in our fit to only include data points of \citet{bouwens14} which are derived from a more comprehensive sample.

The best Schechter fit (solid blue curve in Fig. \ref{fig:lf7}) yields the following parameters: $M^{\star}_{UV} = -20.63^{+0.69}_{-0.56}$ mag, $\phi^{\star} = 4.59\pm0.38 \times10^{-4}$ Mpc$^{-3}$, and $\alpha =-1.88^{+0.17}_{-0.20}$. Together with these results, we also present the 2D probability distribution of the two Schechter parameters $M^{\star}$ and $\alpha$ in Fig. \ref{fig:lh7}. The dark blue and light blue contours represent the 68\% and 95\% confidence intervals, respectively. We also perform a Schechter fit while fixing the parameter $M^{\star}$ to the value of -20.64 mag of \citet{bouwens14} because our data do not extent to magnitudes brighter than $-20$. The best fit parameters remain close to the values derived before with $\phi^{\star} = 4.5\pm0.5 \times10^{-4}$ Mpc$^{-3}$ and $\alpha = -1.89\pm0.06$. As pointed out earlier, galaxies close to the detection limit of the {\em HST} data are very sensitive to the lensing effects, photometric scatter and contamination. Therefore the faintest bin becomes in turn very sensitive to the completeness correction, i.e the effective survey volume. The faintest bin in the $z \sim 7$ UV LF of \citet{schenker13} is a good illustration of this effect, as their data point is much below our new estimate. Therefore caution must be exercised regarding the volume density at $M_{UV} > -16$. For comparison we also performed a Schechter fit excluding the faintest bin, which yields the following parameters: $M^{\star}_{UV} =-20.95^{+0.89}_{-0.71}$ mag, $\phi^{\star} = 2.54\pm3.65 \times10^{-4}$ Mpc$^{-3}$, and $\alpha =-2.03^{+0.19}_{-0.28}$.

\begin{table}
\centering
\caption{Comparison of the best fit $z \sim 7$ LF Parameters}
\label{tab:lf}
\begin{tabular}[c]{l c c c}
\hline
\hline
Reference &  $M^\star$ & $\alpha$ & $\log_{10} \phi^\star$ \\
Reference &  & & [Mpc$^{-3}$] \\
\hline
A2744 (This work)$^{a}$	& $-20.63^{+0.69}_{-0.56}$	& $-1.88^{+0.17}_{-0.20}$	& $-3.34 \pm 0.36$	 \\
A2744 (This work)$^{b}$ 	& $-20.95^{+0.89}_{-0.71}$	& $-2.03^{+0.19}_{-0.28}$	& $-3.58 \pm 0.62$	 \\
\citet{ishigaki14}	& $-20.2 \pm 0.3$	& $-2.10^{+0.30}_{-0.15}$	& $-3.15^{+0.35}_{-0.30}$	 \\
\citet{bouwens14}	& $-21.04 \pm 0.26$	& $-2.06 \pm 0.12$	& $-3.65^{+0.27}_{-0.17}$	 \\
\citet{schenker13}	& $-20.14^{+0.36}_{-0.48}$	& $-1.87^{+0.18}_{-0.17}$	& $-3.19^{+0.27}_{-0.24}$	 \\
\cite{mclure13}        & $-19.90^{+0.23}_{-0.28}$	   & $-1.90^{+0.14}_{-0.15}$ 	& $-3.35^{+0.28}_{-0.45}$ \\ 
\hline
\multicolumn{4}{l}{\textsc{Note.} -- The results of this work include two LF determinations,} \\
\multicolumn{4}{l}{including or not the faintest (most uncertain) magnitude bin} \\
\multicolumn{4}{l}{(cf. text for details)} \\
\multicolumn{4}{l}{ $^\textrm{a}$ Best Schechter fit to all data}\\
\multicolumn{4}{l}{ $^\textrm{b}$ Best Schechter fit excluding the faintest magnitude bin}\\
\end{tabular}
\end{table}

Comparing our UV LF determination with the literature, it is clear from Fig. \ref{fig:lf7} that our results are in good agreement with the results of the deepest blank fields. Despite the uncertainties on the bright-end and the faintest bin of the LF, our best fit parameters are compatible, within the error bars, with those of \citet{schenker13} and \citet{bouwens14}. In particular, our value of $\alpha =-1.88^{+0.17}_{-0.20}$ is close to $-1.87^{+0.18}_{-0.17}$ derived by \citet{schenker13} or $\alpha = -1.9\pm0.14$ by \citet{mclure13}, but lower than $\alpha = -2.01\pm0.14$ of \citet{bouwens14}, or $\alpha = -2.06\pm0.12$ from their analysis of a compilation of {\em HST} fields (see Table \ref{tab:lf}). While the uncertainties associated with the different determinations prevent us from a definitive conclusion regarding the value of $\alpha$ at $z \sim 7$, they generally agree on a steepening of the faint-end compared to lower redshift results \citep{bouwens14}. More recently, \citet{ishigaki14} also carried out an analysis of high-redshift galaxies behind the A2744 cluster. They report a steeper faint-end slope of $-2.10^{+0.30}_{-0.15}$, which goes down to intrinsic magnitudes of $M_{UV}=-17$ mag, clearly shallower than our limit of $M_{UV}=-15.5$ mag. This can be due to several differences observed between their study and ours. First, they report a lower number of dropout candidates, likely due to different source extraction parameters and photometric measurements. Second, they rely on pre-HFF observations to build a mass model, while our model uses the latest set of $\sim 150$ multiple images identified in HFF observations to derive a very precise lensing model. Indeed, the average amplification predicted by the new model of A2744 appears higher than in pre-HFF models \citep{jauzac14}, which might explain the shallower limit of the $z \sim 7$ LF in \citet{ishigaki14}.

Gravitational lensing enables us to reach much fainter galaxies than any blank fields before. We are able to extend the search for high-$z$ galaxies down to an absolute magnitude of $M_{UV} \sim -15.5$ mag. The outcome is that the steep slope of the faint-end still holds below $M_{UV} =-17$ mag, suggesting that ultra-faint galaxies still dominate the galaxy number density at redshifts around $z \sim 7$. This result has important implications for the reionization of the Universe, as dwarf galaxies are thought to be the most likely culprits \citep[e.g.,][]{robertson10}. The ability of faint galaxies to maintain the reionization of the intergalactic medium is deferred to a future study.

\subsection{The UV LF at $z\sim 8$}
\label{sec:lf8}

The luminosity distribution of $z\sim8$ sources consists of three bins at M$_{abs,UV}$=[-20,-19,-17] with a bin width of 1 mag for the two brightest bins and 2 mag for the faintest bin, respectively. The size of the redshift $z \sim 8$ sample, although close to the HFF predictions, is significantly smaller than $z \sim 7$. We note that there is only one candidate in the last magnitude bin which makes the LF determination prone to large uncertainties, which is also true for the other bins due to small number statistics. Nevertheless, it is interesting to compare our results to previous determinations of the $z \sim 8$ UV LF in blank fields.

\begin{figure}[!htbp]
   \centering
   \includegraphics[width=9cm]{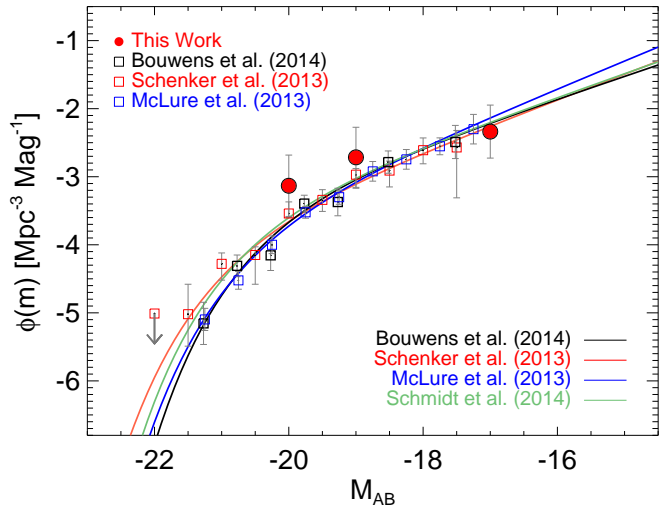} 
     \caption{The UV galaxy luminosity function at $z \sim 8$. The red circles represent our estimate of the UV luminosity distribution based on the effective volume method. For comparison, the results of \citet{bouwens14} are shown with black open squares and their Schechter function is shown with a black dashed line. The red points represent the determination of \citet{schenker13} with a best fit shown with the red dashed line, while the blue open squares represent the results of \citet{mclure13} and the blue dashed line their Schechter fit. We also include data points from the BoRG survey \citet{schmidt14}, which explores the bright-end of the luminosity function without binning in magnitude (green curve).}
   \label{fig:lf8}
\end{figure}

We see in Fig. \ref{fig:lf8} that our LF points are globally higher than recent estimates of \citet{schenker13}, \citet{bouwens14}, and \citet{schmidt14} towards the bright-end. This might be due to the presence of an overdensity of high-$z$ galaxies in the North-East part of the image (cf. Fig\ref{fig:layout}), which was also discussed in \citet{zheng14} and \citet{coe14}. \citet{ishigaki14} observe the same excess in the number counts of $z \sim 8$ objects towards the bright-end of the LF. This indicates that cosmic variance is probably more important than expected at these redshifts as discussed very recently in \citet{robertson14}. The data point on the faintest bin is however in line with the best fit results in the literature, showing that similarly to the LF at $z \sim 7$, there is no evidence of a turnover. The steep slope of the faint-end slope means galaxies as faint as $M_{UV} \sim -17$ mag become more numerous at higher redshift, confirming the steepening of $\alpha$ found in \citet{mclure13} among other studies. Given the uncertainties in play, we cannot yet put strong constraints on the shape of the LF at $z \sim 8$. Yet this result shows the potential of the HFF program in extending this type of studies to very faint magnitudes, and the completion of program with the observations of six galaxy clusters in total will help mitigate such effects thanks to a larger survey area.

\section{Conclusion}
\label{sec:conclusion}

Using the complete {\em HST} dataset of the first Hubble Frontier Fields cluster A2744, we present new constraints on the faint-end of the UV galaxy luminosity function at redshift $z \sim 7$ and $z \sim 8$ in this work. This is the first of a series of six clusters that will ultimately be observed by {\em HST} and other telescopes to study the distant Universe to unprecedented level of depth and detail. Thanks to the boost offered by gravitational lensing of these massive galaxy clusters, one can reach the most feeble background galaxies at high redshift. In the present study, we uncovered the faintest galaxies known to date at $z \sim 7$ and 8, reaching down to absolute magnitudes of $M_{UV} \sim -15.5$. The flux of the background galaxies in our sample is amplified by factors up to 30, while the total survey area is reduced to $\sim0.96$ acrmin$^{2}$ in the source plane.

We have selected a total of 50 $I_{814}-$dropout candidates at $\sim 7$ and eight $Y_{105}-$dropouts at $z \sim 8$ using color-color selection based on the Lyman break technique. Three of these systems are multiply-imaged by the lensing cluster. In order to compute the UV galaxy luminosity function, we have performed comprehensive Monte Carlo simulations to estimate the effective survey volume, including contamination and lensing effects. Thanks to the updated lensing model of A2744 constructed by our CATS team \citep{jauzac14,richard14}, we were able to accurately compute the amplification factor for each galaxy and estimate the completeness correction for each magnitude bin.

We were able to constrain the faint-end slope of the $z \sim 7$ UV LF to unprecedented depth. Our best Schechter fit is in good agreement with different results from blank field observations, the most recent being the UDF12 campaign \citet{schenker13,mclure13} and the {\em HST} legacy fields \citet{bouwens14}. In  particular, we find a faint-end slope of  $\alpha = -1.88^{+0.17}_{-0.20}$ close to the previous determinations. But most interestingly, we were able to put constraints on the number density of the faintest galaxies at $z \sim 7$, showing no strong evidence of a turnover in the steep slope of the UV LF down to $M_{UV} \sim -15.5$. We confirm for the first time that the slope of the LF remains steep down to luminosities about 0.01$L^{\star}$. According to recent cosmological simulations, a faint-end slope $\alpha \sim -1.9$ to $-2.3$ is predicted to remain steep down to luminosities as faint as $M_{UV} \sim -13$ to $-10$ mag \citep[e.g.][]{faucher-giguere11, jaacks12,kimm14}.

Our results at redshift $z \sim 8$ are also in line with blank field estimates. Although the small number statistics due the reduced survey area prevent us from establishing strong constraints on the UV LF, our results show the first hints of a steep slope of the $z\sim 8$ luminosity function at luminosities below 0.1$L^{\star}$.

Overall, our results confirm the ability of strong lensing, and the HFF program in particular, to effectively probe the epoch of reionization. This approach is complementary to the blank fields or wide area surveys that provide better constraints on the brighter part of the luminosity function. While the variety of blank fields provide a large number of high-redshift galaxies, the HFF clusters reach the faintest galaxies, down to an apparent magnitude of m$_{AB} \sim 32$, which would otherwise remain inaccessible with current instrumentation. The completion of the HFF observations of six lensing clusters will help mitigate cosmic variance and uncertainties by increasing the sample size of galaxies lying at $z \gtrsim 7$. We will be able to put even better constraints on the faint-end slope of the UV LF and its evolution at $z \gtrsim 7$, and infer the role of galaxies in the cosmic reionization.

\begin{table*} 
\centering
\caption{\label{tab:photometry7} Photometric and color measurements for the $z \sim 6-7$ dropouts}
\begin{tabular}{lccccccccc}
\hline
Target & R.A. (J2000) & Dec (J2000) & $I_{814}-Y_{105}$ & $Y_{105}-J_{125}$ & $J_{125}$  & Magnification\footnote{this is the flux amplification factor} & Photo$-z$&$z$ low 68\%& $z$ high 68\% \\ \hline    
260    &   3.5938064    &    -30.415444   & 	 2.60     $\pm$   0.61     &   0.03   $\pm$	  0.10	  &    26.38	 $\pm$      0.09 &   7.47  &  6.60& 6.367 & 6.769 \\  
330    &   3.5706506    &    -30.414661   & 	 1.45     $\pm$   0.22     &   0.21   $\pm$	  0.12	  &    26.26	 $\pm$      0.10 &   2.12  &  6.00& 5.730 & 6.210 \\
491    &   3.5929474    &    -30.413330   & 	0.96     $\pm$   0.33      &  -0.11   $\pm$	  0.23	  &    27.87	 $\pm$      0.21 &  16.14  &  5.82& 5.467 & 6.075 \\
800    &   3.6000749    &    -30.411626   & 	0.95     $\pm$   0.58      &  -0.18   $\pm$	  0.31	  &    28.08	 $\pm$      0.28 &  17.31  &  5.68& 4.841 & 6.197 \\
860    &   3.5782841    &    -30.410941   &  $>$    0.91     		   &   0.10	  $\pm$   0.34	   &   28.34	 $\pm$      0.29 &   4.36  &  5.79& 4.978 & 6.173 \\
980    &   3.6030396    &    -30.410559   &  $>$     2.19		  &    0.35   $\pm$	  0.39    &    28.46    $\pm$	    0.30 &   7.27  &  6.70& 3.930 & 7.557 \\
1011   &    3.6006172   &     -30.410297  &  $>$      2.03		  &    0.06   $\pm$	  0.19   &     27.84	$\pm$	    0.16 &  18.21  &  6.37& 6.116 & 6.543  \\
1061   &    3.5923551   &     -30.409892  &  $>$      2.54		  &    0.50  $\pm$	  0.28   &     27.95   $\pm$	    0.21 &  15.25  &  7.19& 6.615 & 7.474  \\
1071   &    3.6032119   &     -30.410353  &  $>$      2.26		  &    0.08   $\pm$	  0.17   &     27.05	$\pm$	    0.14 &   6.95  &  6.42& 5.935 & 6.828 \\  
1118   &    3.6041423   &     -30.409562  &  $>$      2.12		  &   -0.11  $\pm$	  0.51   &     28.99   $\pm$	    0.45 &   5.73  &  6.23& 4.301 & 7.204 \\
1168   &     3.6047795  &      -30.409228 &  $>$       2.56		  &    0.07  $\pm$	  0.28  &      28.36   $\pm$	    0.24 &   5.16  &  6.55& 6.258 & 6.803 \\
1184   &     3.6045663  &      -30.409361 &  $>$       1.13		  &    0.20 $\pm$	  0.30  &      28.11   $\pm$	    0.24 &   5.24  &  5.83& 5.193 & 6.332 \\
1282   &     3.5754838  &      -30.408586 &  $>$       2.19		  &   -0.15   $\pm$	  0.45  &      28.96	 $\pm$      0.40 &   4.12  &  6.40& 5.865 & 6.799 \\
1466   &     3.5960318  &      -30.407581 &  $>$       1.14		  &   -0.31   $\pm$	  0.45  &      28.87	 $\pm$      0.41 &   8.14  &  5.98& 5.464 & 6.391 \\
1467   &     3.5693348  &      -30.407412 &  $>$   1.07		         &    -0.33   $\pm$       0.39  &      29.23	 $\pm$      0.36 &   2.65  &  5.78& 5.008 & 6.184 \\
1569   &     3.6063862  &      -30.407284 &        1.71  $\pm$     0.31  &     0.06 $\pm$	  0.14  &      26.79   $\pm$	    0.12 &   3.94  &  6.28& 6.120 & 6.435 \\
1894   &     3.6009451  &      -30.405598 &  $>$   2.24		          &    0.13 $\pm$         0.35  &      28.63    $\pm$       0.29 &  10.93  &  6.81& 4.582 & 7.187 \\
2073   &     3.5804527  &      -30.405042 &  $>$   2.49		          &    0.26 $\pm$         0.13  &      26.80     $\pm$      0.10 &  11.06  &  6.96& 6.481 & 7.117 \\
2223   &     3.5991119  &      -30.404088 &  $>$   2.19		          &    0.09  $\pm$	  0.45  &      28.72	 $\pm$	    0.37 &  15.47  &  6.57& 5.803 & 7.084 \\
2270   &     3.6010706  &      -30.403993 & 	   1.33  $\pm$     0.45  &     0.16      $\pm$	  0.18  &      27.51   $\pm$	    0.15 &   8.32  &  6.00& 5.529 & 6.291 \\
2391   &     3.5746883  &      -30.403316 &  $>$   2.19			  &   -0.09  $\pm$        0.43	 &     28.90	  $\pm$     0.38 &   4.89  &  6.38& 5.709 & 7.052  \\
2800   &     3.5798438  &      -30.401593 &  $>$   2.21			  &    0.31  $\pm$        0.30	 &     28.47	  $\pm$     0.23 &  17.78  &  6.79& 4.157 & 7.442  \\
2871   &     3.5874513  &      -30.401374 & 	   1.52  $\pm$     0.37  &     0.01  $\pm$        0.14	 &     27.69  $\pm$	    0.12 &   6.28  &  6.11& 5.809 & 6.335 \\
2884   &     3.5677694  &      -30.401279 &  $>$   2.96			 &    -0.09  $\pm$	  0.36  &      28.14  $\pm$	    0.32 &   2.85  &  6.42& 5.927 & 6.796 \\
3030   &     3.5976037  &      -30.400430 &  $>$   1.25			 &    -0.16 $\pm$	  0.34  &      28.78  $\pm$	    0.30 &  12.08  &  6.05& 5.715 & 6.391 \\
3294   &     3.5893501  &      -30.398708 &  $>$   1.08			 &     0.01  $\pm$	  0.42  &      28.81  $\pm$	    0.36 &  19.03  &  6.00& 5.235 & 6.575 \\
3368   &     3.5734563  &      -30.398395 &  $>$   0.96			  &   -0.21 $\pm$	  0.47  &      28.79  $\pm$	    0.42 &   7.70  &  5.91& 5.298 & 6.311 \\
3437   &     3.5853217  &      -30.397958 & 	   2.10  $\pm$     0.49  &     0.08  $\pm$	  0.11  &      26.99  $\pm$	    0.09 &   8.29  &  6.32& 6.063 & 6.540 \\
3453   &     3.6080366  &      -30.397735 &  $>$   2.98			&      0.06  $\pm$	  0.34  &      27.96  $\pm$	    0.30 &   2.79  &  6.58& 6.284 & 6.883 \\
3772   &     3.5978343  &      -30.395960 &  $>$   2.27			&      0.30 $\pm$	  0.15  &      27.07  $\pm$	    0.12 &   6.82  &  7.03& 6.536 & 7.183 \\
4358   &     3.5862495  &      -30.392708 &  $>$   1.36			&     -0.10 $\pm$	  0.38  &      28.68  $\pm$	    0.34 &  29.85  &  6.08& 5.469 & 6.629 \\
4581   &     3.5703274  &      -30.391252 &  $>$   2.43			&      0.06  $\pm$	  0.44  &      28.52  $\pm$	    0.38 &   3.76  &  6.35& 4.888 & 7.166 \\
4656   &     3.5766562  &      -30.391368 & 	   1.62  $\pm$    0.05  &      0.33 $\pm$         0.01   &     24.00  $\pm$         0.01 &   9.97  &  5.61& 5.497 & 6.037 \\
4671   &     3.5968919  &      -30.390454 &  $>$   2.28			&      0.03  $\pm$        0.41   &     28.69  $\pm$         0.35 &   5.36  &  6.46& 6.202 & 6.739 \\
4779   &     3.6034180  &      -30.383218 & 	   0.88   $\pm$    0.28  &    -0.11 $\pm$	  0.21  &      27.67  $\pm$	    0.19 &   2.07  &  5.79& 5.474 & 6.015 \\
4867   &     3.5770445  &      -30.382586 &  $>$   1.42		      &        0.28  $\pm$        0.28    &    27.97  $\pm$	    0.22 &   3.53  &  6.04& 5.278 & 7.103 \\
5020   &     3.6089971  &      -30.385283 & 	   1.18  $\pm$     0.46  &    -0.03  $\pm$	  0.26  &      28.02  $\pm$	    0.22 &   1.86  &  6.00& 5.645 & 6.274 \\
5038   &     3.6087401  &      -30.384138 &  $>$   2.24			&      0.29 $\pm$	  0.42  &      28.47  $\pm$	    0.33 &   1.89  &  6.30& 5.258 & 7.134 \\
5160   &     3.6062254  &      -30.386647 & 	   1.02  $\pm$     0.18  &     0.15 $\pm$         0.09   &     26.36  $\pm$         0.08 &   2.10  &  5.49& 5.106 & 5.838 \\
5258   &     3.5768893  &      -30.386323 &  $>$   2.96			 &     0.08  $\pm$	  0.24  &      27.96  $\pm$         0.20 &   5.16  &  6.44& 5.928 & 6.723  \\
5398   &     3.5710763  &      -30.386138 & 	   0.89   $\pm$    0.19  &    -0.02  $\pm$	  0.13  &      26.09  $\pm$	    0.12 &   3.69  &  5.80& 5.485 & 5.964 \\
5798   &     3.5914765  &      -30.390200 &  $>$   1.34			&      0.32 $\pm$	  0.35  &      28.23  $\pm$	    0.27 &   8.12  &  6.08& 5.445 & 6.840 \\
6245   &     3.5905165  &      -30.379759 &  $>$   1.30			&     -0.32 $\pm$	  0.40  &      28.46  $\pm$	    0.37 &   2.73  &  6.10& 5.517 & 6.452  \\
6498   &     3.6065771  &      -30.380923 &  $>$   2.57			&      0.72 $\pm$	  0.33  &      27.70  $\pm$	    0.20 &   1.76  &  5.70& 4.877 & 7.791  \\
6520   &     3.5797768  &      -30.381136 &  $>$   2.35			&     -0.44 $\pm$	  0.50  &      29.10  $\pm$	    0.46 &   3.34  &  6.41& 6.196 & 6.628  \\
6525   &     3.6064597  &      -30.380995 &  $>$   3.54			&      0.73 $\pm$	  0.20  &      26.72  $\pm$	    0.12 &   1.78  &  7.30& 6.822 & 7.735  \\
6595   &     3.5946240  &      -30.380403 &  $>$   1.32			&     -0.04  $\pm$	  0.41  &      28.17  $\pm$	    0.36 &   2.47  &  6.15& 5.704 & 6.602  \\
6601   &     3.5945418  &      -30.381428 &  $>$   2.39			&      0.03  $\pm$	  0.34  &      28.57  $\pm$	    0.29 &   2.91  &  6.44& 6.225 & 6.647  \\
6662   &     3.5981042  &      -30.382389 &  $>$   2.19			&      0.52 $\pm$	  0.39  &      28.28  $\pm$	    0.28 &   2.45  &  7.39& 5.273 & 7.792  \\
6670   &     3.5867218  &      -30.381426 &  $>$   0.83			  &    0.01  $\pm$	  0.47  &      28.84  $\pm$	    0.41 &   3.38  &  5.71& 4.860 & 6.276  \\ \hline
\end{tabular}
\end{table*}

\begin{table*}
\centering
\caption{\label{tab:photometry8} Photometric and color measurements for the $z \sim 8$ dropouts}
\begin{tabular}{lccccccccc}
\hline
Target & R.A. (J2000) & Dec (J2000) & $Y_{105}-J_{125}$ & $J_{125}-H_{140}$ & $H_{140}$  & Magnification\footnote{this is the flux amplification factor} & Photo$-z$ &$z$ low 68\%& $z$ high 68\% \\ \hline    
1060    &     3.5921620    & 	-30.409916   &     0.59   $\pm$    0.38   &	-0.09 $\pm$	0.34	&     28.46  $\pm$    0.22 &16.50& 7.39 &  6.653 & 7.736   \\
4877    &     3.6033808    & 	-30.382255   &      1.06   $\pm$   0.20   &	 0.24 $\pm$	0.12	&     26.38  $\pm$    0.06 & 2.03& 7.68 &  7.242 & 7.979   \\
4886    &     3.6038552    & 	-30.382263   &      1.44   $\pm$   0.10   &	 0.22 $\pm$	0.17	&     26.98  $\pm$    0.09 & 2.00& 7.99 &  7.135 & 8.317   \\
5202    &     3.5960933    & 	-30.385831   &      1.20   $\pm$   0.22   &	 0.12 $\pm$ 	0.13	&     26.69  $\pm$    0.07 & 3.41& 8.00 &  7.644 & 8.243   \\
5918    &     3.5951375    & 	-30.381131   &      1.16   $\pm$   0.13   &	 0.12 $\pm$     0.08	 &    26.79  $\pm$    0.05 & 3.47& 7.71 &  7.408 & 8.043   \\
6436    &     3.6045192    & 	-30.380465   &      1.05   $\pm$   0.18   &	 0.09  $\pm$    0.12	&     25.99  $\pm$    0.07 & 1.85& 7.70 &  7.253 & 7.963   \\
6504    &     3.5889794    & 	-30.378665   &      1.24   $\pm$   0.48   &     -0.07  $\pm$    0.28	&     28.07  $\pm$    0.19 & 2.68& 7.89 &  6.855 & 8.231   \\
6593    &     3.6050570    & 	-30.381462   &      1.15   $\pm$   0.46   &	-0.39 $\pm$	0.33	&     28.44  $\pm$    0.25 & 1.88& 7.84 &  7.155 & 8.125   \\ \hline
\end{tabular}
\end{table*}

\begin{table*}
\centering
\caption{\label{tab:multiples} Multiply-imaged systems at $z \sim 6-9$}
\begin{tabular}{lccccccc}
\hline
Image & Target & R.A. (J2000) & Dec (J2000) & $I_{814}-Y_{105}$ & $Y_{105}-J_{125}$ & $J_{125}-H_{140}$ & $H_{140}$   \\ \hline 
1.1 & 1061 & 3.592355 & -30.409892   &   $>$ 2.54                   &    0.51    $\pm$   0.29 &    -0.27   $\pm$    0.29  &   28.22   $\pm$    0.21 \\
1.2 & 2487 & 3.588287 & -30.410333   &  0.66    $\pm$   1.45    &  0.56       $\pm$ 0.90   &   0.07    $\pm$   0.78    & 29.32    $\pm$   0.46  \\
1.3 & 2978 & 3.600937 & -30.400809   &  1.09    $\pm$   2.32    &  0.88       $\pm$ 0.66   &   0.19    $\pm$   0.46    & 28.83    $\pm$   0.25   \\
2.1 & 2800 & 3.579844 & -30.401593   & $>$  2.21                    & 0.32         $\pm$  0.30  &    0.09  $\pm$     0.29  &   28.38   $\pm$    0.17 \\
2.2 & 8170 & 3.583566 & -30.396720      &$>$   1.00                    &     0.46     $\pm$  0.60  &    0.04    $\pm$   0.55  &   29.50    $\pm$   0.33 \\
5.1 & 2073 &     3.5804527   &   -30.405042   &   $>$  2.49                           &  0.26      $\pm$   0.13   &  -0.16    $\pm$   0.14    &   26.96   $\pm$   0.09  \\   
5.2 & 2871 &     3.5874513    &  -30.401374   &    1.52    $\pm$  0.37            &  0.01  $\pm$     0.15     &   0.02   $\pm$    0.16  &    27.67  $\pm$    0.09   \\ 
5.3 & 3437 &     3.5853217    &  -30.397958   &    2.10   $\pm$   0.49            &  0.08   $\pm$    0.11     & -0.16     $\pm$  0.13     &  27.15    $\pm$  0.08   \\
5.4 & 3772 &     3.5978343   &   -30.395960   &  $>$   2.27                           &   0.30     $\pm$ 0.15     & -0.12     $\pm$   0.16    &   27.19    $\pm$   0.10  \\ \hline
\end{tabular}
\end{table*}

\acknowledgments

We want to thank the STScI and the HFF team for their efforts in obtaining and reducing the {\em HST} data. HA and JPK are supported by the European Research Council (ERC) advanced grant ``Light on the Dark'' (LIDA). JR acknowledges support from the ERC starting grant CALENDS. MJ acknowledges support from the Leverhulme Trust (grant number PLP-2011-003) and Science and Technology Facilities Council (grant number ST/L00075X/1). PN acknowledges support from NSF theory grant AST-1044455 and a theory grant from Space Telescope Science Institute HST-AR1214401.A. ML acknowledges support from CNRS. 

\bibliographystyle{apj}
\bibliography{references.bib}

\end{document}